# Boltzmann-Fokker-Planck Kinetic Solver with Adaptive Mesh in Phase Space

Vladimir Kolobov,[1,2 a)] Robert Arslanbekov[1] and Dmitry Levko[1]

[1]*CFD Research Corporation, 701 McMillian Way, Huntsville, AL 35806, USA.*
[2]*The University of Alabama in Huntsville.*

[a)]Corresponding author: vik@cfdrc.com

**Abstract.** We describe the implementation of kinetic solvers in 1d2v phase space using adaptive Cartesian mesh. Spherical coordinates in velocity space are used to simplify the Lorentz and Fokker-Planck collisional operators. The key capabilities of the new solvers are illustrated for electron elastic scattering, acceleration, continuous energy loss in collisions, and ionization processes in weakly-ionized plasma. We have also implemented two-stream approach to reduce computational cost for studies of gas breakdown dynamics in the presence of runaway electrons. The benefits and limitations of the non-split-phase-space method for kinetic solvers with adaptive mesh in phase space are discussed.

## INTRODUCTION

Simulation of particle kinetics in weakly-ionized plasma remains a challenging task in computational physics. With increasing computational power, kinetic solvers based on discrete velocity method (DVM) become competitive with the particle-based solvers.[1,2] We have previously demonstrated an Adaptive Mesh in Phase Space (AMPS) method for reducing computational cost of the DVM kinetic solvers.[3] Our AMPS solvers were based on splitting velocity and configuration (physical) space. The present paper is devoted to another strategy, i.e. utilizing symmetry of the problem to reduce dimensionality of kinetic solvers. We describe the implementation and application of a Boltzmann-Fokker-Planck solver using adaptive Cartesian mesh in 2d and 3d phase space without splitting velocity and configuration spaces.

Using AMPS technique without splitting phase space transport creates challenges for coupling kinetic solvers with electromagnetics. Kinetic solvers operate in the phase-space of higher dimensionality compared to the electromagnetic solvers operating in physical space. When using AMPS without splitting, mesh adaptation in velocity space triggers mesh adaptation in physical space and vice versa. Calculation of particle density and flux required for the electromagnetic solvers involves integration of the velocity distribution function (VDF) over velocity space. We have previously developed accurate interpolation techniques to couple 1d2v Vlasov and Fokker-Planck solvers with 1d Poisson solver.[4] In the present paper, we extend this technique to the Boltzmann-Fokker-Planck solvers for electrons in weakly-ionized collisional plasma.

Collisions of electrons with neutrals and Coulomb interactions among charged particles are best described using spherical coordinates in velocity space. For transport in configuration (physical) space, Cartesian, cylindrical or spherical coordinate systems can be used depending on problem type. Electron collisions with neutrals can be divided into two types. Boltzmann-type collisions are associated with a large change of electron momentum – they are described by an integral operator in velocity space. Collisions of the Fokker-Planck-type are associated with small changes of momentum or energy. They are described by Fokker-Planck-type (drift-diffusion) differential operators in velocity space. We solve Boltzmann-Fokker-Planck (BFP) kinetic equation using Finite Volume method with octree Cartesian mesh, without splitting physical and velocity spaces.

We demonstrate our BFP solver for simulation of electron scattering in elastic collisions, acceleration by electric field, continuous energy loss in collisions, and generation of secondary electrons by ionization. We analyse the conditions for generation of runaway electrons in a spatially inhomogeneous electric field and illustrate effects of scattering on the runaway effect. We implement two-stream approach to reduce the computational cost of kinetic simulations. By coupling BFP and Poisson solvers, we study the dynamics of nanosecond gas breakdown by electric pulses in the presence of runaway electrons. For studies of processes developing on (fast) electron time scale, ions are assumed motionless.

# BOLTZMANN-FOKKER-PLANCK KINETIC EQUATION FOR ELECTRONS

The Boltzmann-Fokker-Planck kinetic equation for electrons in partially-ionized, non-magnetized plasma has the form:

$$\frac{\partial f}{\partial t} + \nabla \cdot (\boldsymbol{v}f) + \frac{1}{m}\frac{\partial}{\partial v}\left[\left(F(v)\frac{v}{v} - e\boldsymbol{E}\right)f\right] + \frac{\partial}{\partial v_i}\left[D_{ij}\frac{\partial f}{\partial v_i}\right] = I_c. \tag{1}$$

The left-hand side of Eq. (1) describes electron acceleration by the electric field $\boldsymbol{E}$ and collisional processes associated with small changes of energy and momentum written in the Fokker-Planck form. The friction force, $F(v)\frac{v}{v}$, is directed towards zero velocity. Small-angle scattering is described by a diffusion tensor

$$D_{ij} = \nu_e(v)v^2\left(\delta_{ij} - \frac{v_i v_j}{v^2}\right) \tag{2}$$

where $\nu_e(v)$ is the transport collision frequency. The right-hand side of Eq. (1) describes Boltzmann-type collisions associated with large changes of energy and momentum. For example, elastic scattering of electrons with neutral particles can be described by Lorentz model: [5]

$$I_e = \nu_c(v)\left[\int_{S^2} p(\Theta)f(v)d\Omega - f\right], \tag{3}$$

where $\nu_c(v)$ is the total collision frequency, and $p(\Theta)$ is a phase function giving the probability of electron scattering with an angle $\Theta$ between the initial and final velocity. The integration in Eq. (3) is performed over a sphere $S^2$ with a radius $v$ in velocity space. Inelastic collisions can be added to the right-hand side of Eq. (1) or included in the friction force (at high electron energies), as described for example by Strickland *et al*.[5]

## Spherical Coordinates in Velocity Space

Here, we will consider one-dimensional, planar problems and will use spherical coordinates in velocity space. Assuming that the axis in velocity space is directed along the direction of the electric force, and the distribution function does not depend on the azimuthal angle $\varphi$, we can rewrite Eq. (1) in the form:

$$\frac{\partial f}{\partial t} + v\mu\frac{\partial f}{\partial x} - \frac{eE}{m}\left(\mu\frac{\partial f}{\partial v} + \frac{1-\mu^2}{v}\frac{\partial f}{\partial \mu}\right) = \frac{1}{v^2}\frac{\partial}{\partial v}(v^2 Ff) + D_\mu\frac{\partial}{\partial \mu}\left((1-\mu^2)\frac{\partial f}{\partial \mu}\right) + I_e, \tag{4}$$

where $\mu = \cos\theta$, and $\theta$ is the pitch angle. Now, the diffusion tensor (2) has only one component, $D_\mu(v)$, and the Lorentz operator (3) involves one-dimensional integral:

$$I_e(v,\mu) = \nu_e(v)\left[\int_{-1}^{1} p(\mu,\mu')f(\mu')d\mu' - f(v,\mu)\right], \tag{5}$$

where $p(\mu,\mu') = \int_0^{2\pi} p(\cos\Theta)\,d\varphi$ is the azimuthally integrated phase function, and $\cos\Theta = \mu\mu' + \left((1-\mu^2)(1-\mu'^2)\right)^{1/2}\cos(\varphi - \varphi')$. Due to the symmetry, the phase function obeys $p(-\mu,\mu') = p(\mu,-\mu')$, $p(-\mu,-\mu') = p(\mu,\mu')$, and is equal to ½ for isotropic scattering.

Introducing $Y = v^2 f$, Eq. (4) can be rewritten in the conservative form:

$$\frac{\partial Y}{\partial t} + \frac{\partial}{\partial x}(UY) + \frac{\partial}{\partial v}(VY) + \frac{\partial}{\partial \mu}[WY] = \frac{\partial}{\partial \mu}\left[D_\mu \frac{\partial Y}{\partial \mu}\right] + v^2 I_c \qquad (6)$$

where

$$U = \mu v; \quad V = eE\mu - F; \quad W = eE\frac{(1-\mu^2)}{mv}.$$

Equation (6) was solved numerically in 1d2v phase space $(x, v, \mu)$ using Basilisk – an open-source framework for solving partial differential equations on adaptive Cartesian meshes. [6]

## BFP SOLVER AND ITS APPLICATIONS

In this section, we demonstrate the basic capabilities of the BFP solver for simulation of electron scattering, acceleration by an electric field, and energy loss in inelastic collisions. We then introduce a two-stream model to analyze the formation of runaway electrons in spatially inhomogeneous electric field in 1d1v phase space. Finally, by coupling the BFP and Poisson solvers, we simulate dynamics of nanosecond gas breakdown by electric pulses in the presence of runaway electrons.

### Elastic Scattering with Arbitrary Anisotropy

We first consider the problem of scattering, which is important for radiation transport and particle kinetics in space and laboratory plasmas. A simple kinetic equation in a two-dimensional $(x, \mu)$ phase space is often used for cosmic rays:[7,8]

$$\frac{\partial f}{\partial t} + v\mu\frac{\partial f}{\partial x} + \frac{v}{2L}\frac{\partial}{\partial \mu}\left((1-\mu^2)f\right) + vf = v\int_{-1}^{1} p(\mu, \mu')f(\mu')d\mu' \qquad (7)$$

where the third term describes the adiabatic focusing in the presence of a diverging magnetic field, and $L(x)$ denotes the spatially-dependent focusing length of the magnetic field. Below, we assume no focusing ($L = \infty$) and use the Rutherford phase function $p(\cos\Theta) = 2\eta(1+\eta)/(1+2\eta-\cos\Theta)^2$ introduced by Strickland et al.[5] to illustrate the influence of scattering anisotropy on the formation of VDF. Figure 1 shows the phase function for two values of the screening parameter $\eta$, which depends on kinetic energy. It is seen that elastic scattering is strongly peaked in the forward direction for $\eta = 0.001$, and becomes close to isotropic at $\eta = 0.1$.

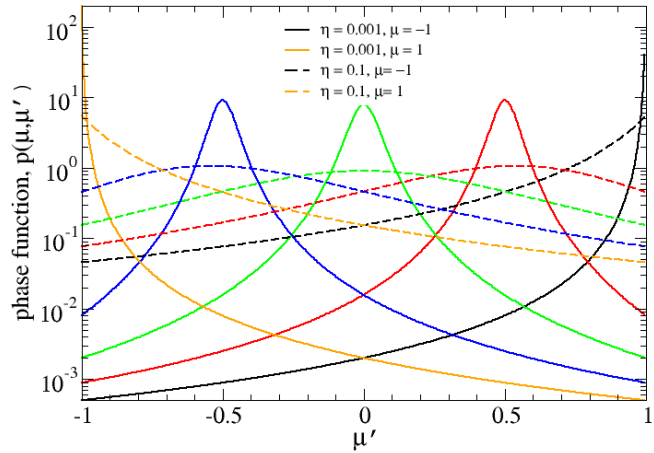

Figure 1. Phase functions for Rutherford scattering at two values of the screening parameter.

Figure 2 shows the results of calculations using the Lorentz and Fokker-Planck collision models. The particles are injected at x=0 and absorbed at x=L. We measure spatial distance in units of scattering length, $\lambda = N\sigma(w)$ where $\sigma(w)$ is the total collision cross-section, and $N$ is the density of scatters. It is seen that the VDF isotropization occurs faster as the scattering becomes more isotropic with increasing $\eta$. This is expected because the isotropization length depends on the transport collision frequency rather than the total collision frequency. Notable differences remain between the diffusion approximation (Fokker-Plank model) and the Lorentz model even at *η = 0.001*.

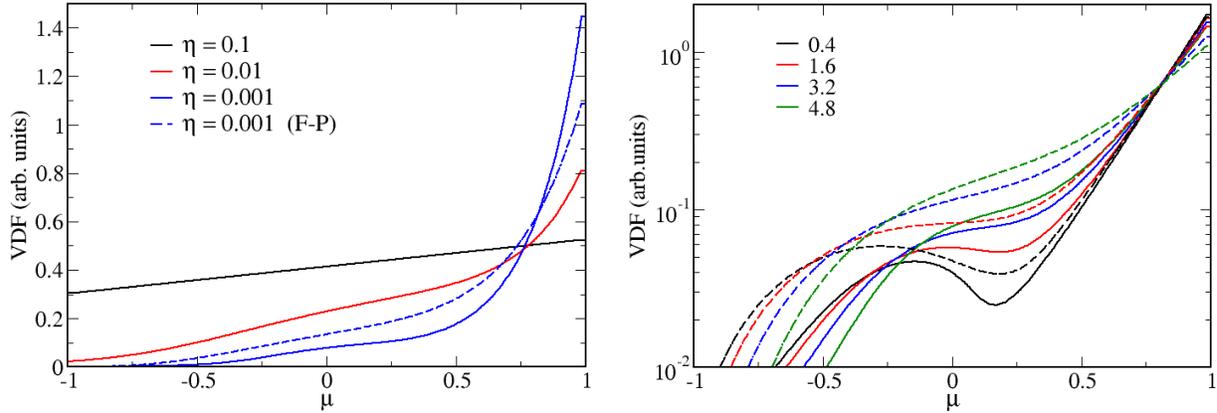

Figure 2. Calculated angular dependencies of VDFs $x/\lambda = 5$ for different screening parameters $\eta$ (on the left), and at different spatial positions, $x/\lambda$ at $\eta = 0.001$ (on the right). Dashed lines show solutions of the F-P equation for the same conditions.

Figure 3 shows axial distributions of electron density and the mean velocity (assuming velocity $v = 1$). We have observed that the total flux (integral over $\mu$) is indeed conserved with good accuracy in our simulations when a steady state is reached. It is seen that at $\eta = 0.1$, the calculated distributions are close to those expected for a linear, diffusion-like solution. The observed time necessary to reach the steady state also agrees with the characteristic diffusion time.

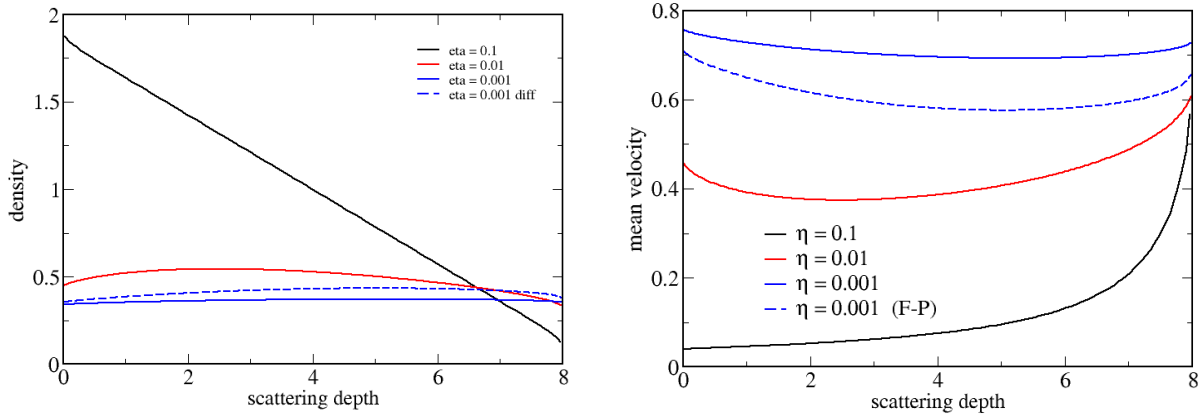

Figure 3. Calculated axial dependencies of electron density and mean velocity (on the right) for different values of the screening parameters $\eta$. Dashed lines show solutions of the F-P equation for the same conditions.

## Electron Acceleration, Scattering, and Energy Loss in Inelastic Collisions

Figure 4 shows the solution of a 1d2v problem of electron acceleration in spatially inhomogeneous electric field, elastic scattering in collisions with neutrals and energy loss in inelastic collisions. Electrons are injected at x=0 and absorbed at $x = L$. The electric field decreases linearly for $x < d$, and is zero at $d < x < L$. We assume a constant loss function, $F_0$, and $eE/F_0 = 5$. Scattering is assumed isotropic, and the electron mean free path is $\lambda = 0.01L$. The VDF is shown at different spatial positions in $(v_\parallel, v_\perp)$ plane. The electron deceleration length, $\Lambda = U_c/F_0$, where $U_c$ is the potential drop. The results resemble those described in Ref [3] for a 1d3v problem, but now obtained as a 1d2v solution at a fraction of computational cost compared to 1d3v computations.

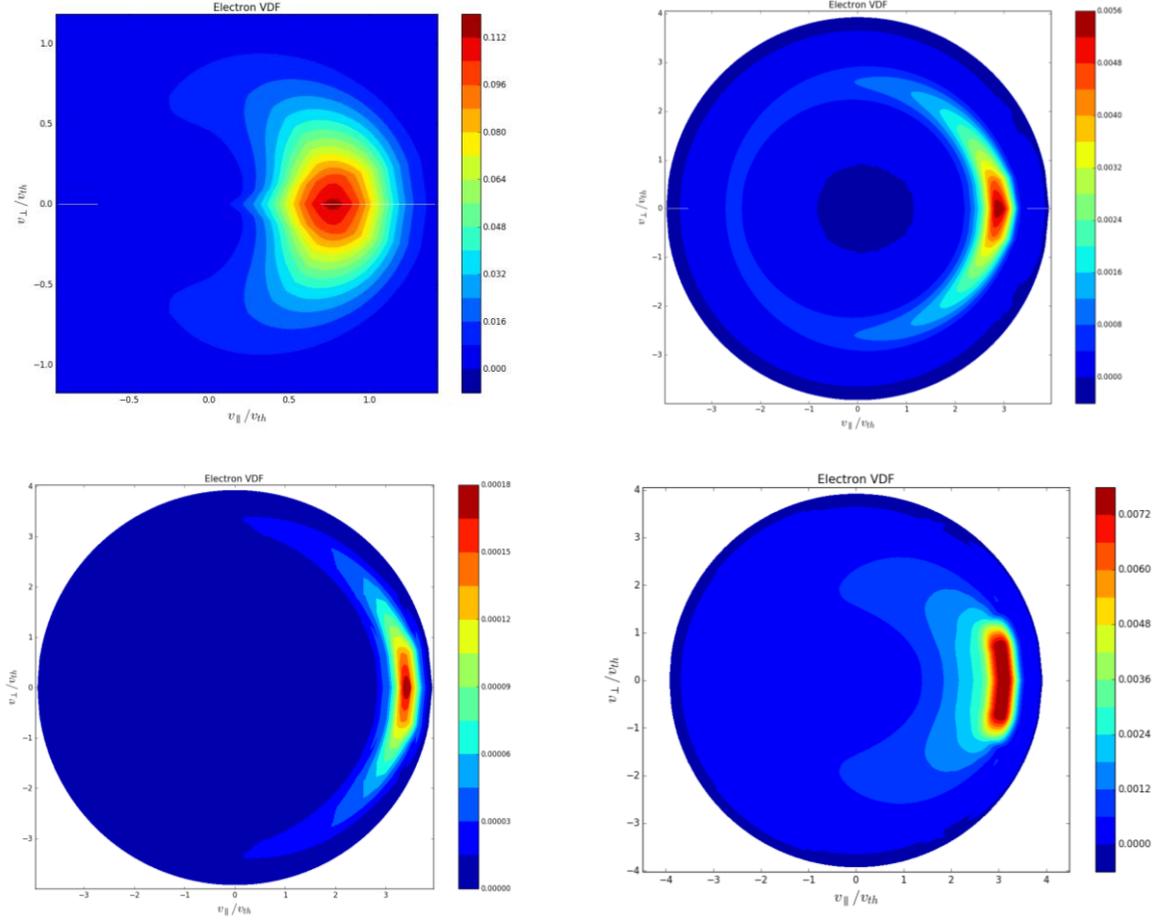

Figure 4. Electron VDF, $Y = v^2 f$, at different spatial positions, $x/L$ = 0.1, 0.2. 0.4. 0.8, for $d/L$ =0.4 and $\lambda = 0.01L$.

## Two-Stream Model

The two-stream model is obtained from the transport equation (4) by integrating it over the forward and backward hemispheres,[9] which corresponds to electrons moving with positive and negative velocities. The Lorentz collision integral becomes

$$I_e(v) = \nu_e(v)\beta(v)[f(-v) - f(v)] \qquad (8)$$

where $\beta$ is the probability that an electron upon elastic scattering goes from one hemisphere to the other: $\beta = 0$ corresponds to complete forward scattering (and hence no coupling between the two streams), $\beta = 1$ implies complete backward scattering, and $\beta = 1/2$ implies isotropic scattering. The general expression for the backscatter ratio can be obtained from the phase function. Stamnes[9] proposed a simple formula for a smooth transition from the forward scattering at high electron energies (described by the Rutherford phase function) to isotropic scattering at low energies:

$$\beta(v) = \begin{cases} \sqrt{\eta(1+\eta)} - \eta & v > v_0 \\ 1/2 & v < v_0 \end{cases} \qquad (9)$$

where $\eta(v) = 0.5(v/v_0)^{-3/2}/(1 - (v/v_0)^{-3/2})$. For Nitrogen, the value of $v_0$ corresponds to kinetic energy of 12 eV. The two-stream model has also been used for simulations of fast electron kinetics in gas discharges.[10] Our approach allows better representation of electron kinetics in 1d1v, as we use electron velocity rather than kinetic energy to take into account particle reflection by the electric field when they move from negative to positive velocities and vice versa. The generation of secondary electron by electron-impact ionization of neutral particles will be described in the simplest form:[11]

$$I_i = \frac{\delta(v)}{\varepsilon_0} \int F(v) f v dv, \qquad (10)$$

where $\delta(v)$ is a velocity distribution of the secondary electrons, and $\varepsilon_0$ is the ionization cost. The velocity distribution of the secondary electrons is assumed to be $\delta(v) = \sqrt{\xi/\pi} \exp(-\xi v^2)$, which at $\xi \to \infty$ becomes a delta-function considered in the analytical model.[11] The ionization frequency $v_i \sim Fv/\varepsilon_0$ is usually smaller than the total frequency of elastic collisions, $v_e(v)$, but the ionization length, which is inverse of the first Townsend coefficient, $\alpha_0 = F_0/\varepsilon_0$, is smaller than the electron deceleration range $\Lambda = U_c/F_0$.

## Electron Runaway in Spatially Inhomogeneous Electric Fields

Electron runaway in spatially inhomogeneous electric field is a complicated process involving acceleration, scattering, energy loss in collisions, and ionization. Here, we consider a 1d1v case using the two-stream model. We assume $F(v) = const = F_0$, and consider a linearly decreasing electric field, $E(x)$, with an amplitude that guaranties the generation of runaway electrons at $eE(x)/F_0 > 1$. The backscattering coefficient is taken from Eq. (9). Figure 5 shows the calculated VDF with account for all the processes mentioned above. Electrons are injected at the left boundary ($x = 0$) with a half-Maxwellian distribution, move along characteristics (stream lines in the left part of Figure 5), and get absorbed at the anode, at $x = L$. Elastic scattering produces jumps between positive and negative values of v. Due to the decrease of the back-scattering rate with electron energy, there is an asymmetry between the positive and negative v: slow electrons have near-isotropic VDF, whereas runaway electrons have strongly anisotropic VDF.

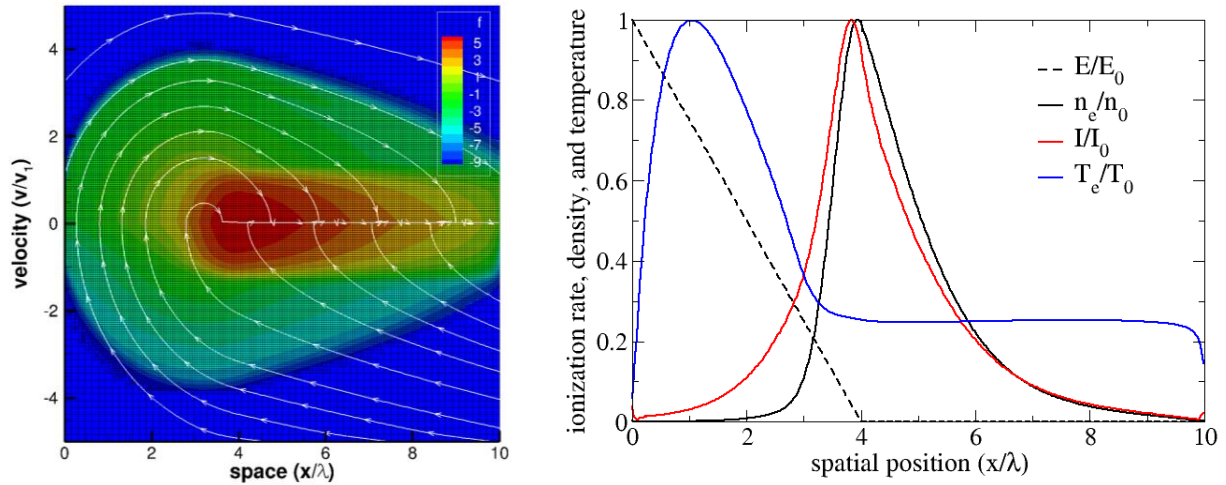

Figure 5. Streamlines (characteristics), adapted computational mesh, and calculated VDF (colour map in log scale) in the 1d1v phase space (left) and spatial distributions of normalized macro-parameters (right).

The right part of Figure 5 shows the spatial distributions of the ionization rate, electron density and temperature, which correspond to the typical distributions observed in the cathode region of short glow discharges without positive column (see Ref. [11] for details).

## Nanosecond Gas Breakdown

Here we demonstrate coupled BFP-Poisson solver for simulations of nanosecond gas breakdown by electric pulses in the presence of runaway electrons. We used the two-stream model, and Eq. (10) for the electron-impact ionization. The friction force $F(v)$ was defined by the Bethe-Bloch law:

$$F(v) = \frac{e^4 ZN}{8\pi\varepsilon_0^2 w} \ln\left(\frac{2w}{I}\right), \qquad (11)$$

where $Z$ is the number of electrons in an atom or molecule, $N$ is the gas density, $\varepsilon_0$ is the permittivity of free space, $w$ is the electron energy, and $I$ is the average energy of inelastic losses. In the absence of scattering, the critical electric field necessary for the electron runaway is

$$E_{cr} = \frac{e^3 ZN}{4\pi\varepsilon_0^2 2.72 I} \qquad (12)$$

For molecular nitrogen ($I = 80$ eV) at pressure of $10^4$ Pa, the critical field is $E_{cr} \approx 45$ kV/cm.

In order to check the prediction of our model, we carried out simulations for two values of the electric field: one of which exceeds the critical electric field (46 kV/cm) and another one is smaller than $E_{cr}$ (36 kV/cm). The time evolution of the VDF in both cases is shown in Figure 6. The VDF shape is mainly defined by the electrons generated due to gas ionization. One can see the presence of runaway electrons for $E > E_{cr}$ and no runaway electrons for $E < E_{cr}$. The local peak of the VDFs at $v_0 \approx 4.5 \times 10^6$ m/s corresponds to the threshold energy $w = I/2$ in Eq. (11) below which the electron energy losses due to the inelastic collisions are zero.

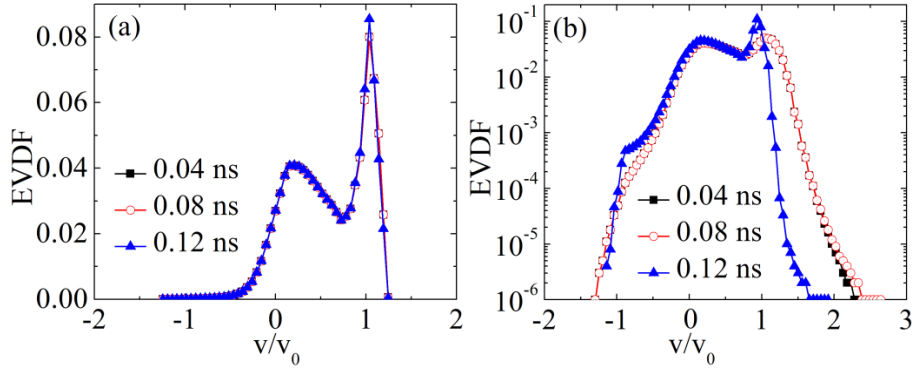

Figure 6. Electron velocity distribution function (in s/m$^4$) obtained in the electric field (a) 36 kV/cm and (b) 46 kV/cm. Gas pressure is $10^4$ Pa.

Now, let us consider the nanosecond time scale breakdown. One can consider two mechanisms of the discharge initiation. In the first mechanism, the discharge is ignited through the multiplication of the initially seeded low-density quasi-neutral plasma ($10^{10}$ m$^{-3}$). The cathode-anode gap is 5 mm. The right electrode (anode) is grounded, while the potential of the left electrode (cathode) increases with time as $dV/dt = 8.4 \times 10^{11}$ V/s. The background gas pressure is $10^4$ Pa. Similar simulation was carried out in Ref. [12] using fluid model and assuming that the ionization rate is defined by the local value of the electric field. The time evolution of the electron number density is shown in Figure 7(a) and the electric potential is shown in Figure 7(b). The time evolution of the VDF in the center of the cathode-anode gap is shown in Figure 7(c).

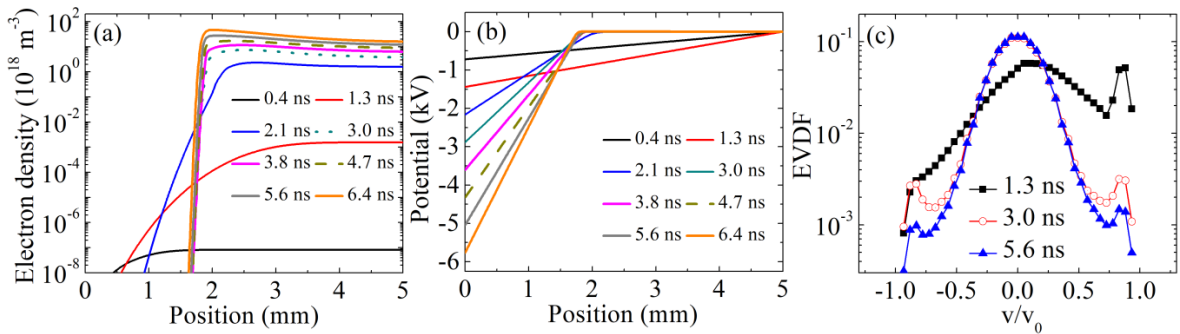

Figure 7. Profiles of (a) the electron density, and (b) the electrostatic potential obtained at different times. (c) Electron velocity distribution function (in s/m$^4$) obtained at the center of the cathode-anode gap at different times.

One can see the formation of the cathode sheath when the plasma density reaches the value ~$10^{18}$ m$^{-3}$. As a result, the highest electric field is obtained in the vicinity of the cathode while the electric field is in the bulk of quasi-neutral plasma is much smaller. The time evolution of the VDF (see Figure 7(c)) shows the local peaks at $v \approx \pm 4.5 \times 10^6$ m/s which corresponds to the threshold of force (11). The electrons belonging to this part of the VDF are responsible for the plasma generation. Figure 7(c) allows one to conclude that there is no the generation of runaway electrons at the given conditions. This is explained by the generation of the dense plasma by the field smaller than $E_{cr}$. One obtains $E > E_{cr}$ only in the cathode sheath which does not contain any electrons which could run away.

In the second mechanism, the discharge is ignited by the electrons emitted from the left electrode into the plasma-free gap. This model can be used to describe the discharges driven by different emission mechanisms such field- or thermo-electron emission, or by electron beams generated by an external source.[13] The cathode-anode gap is 1 cm, the background gas pressure is $10^4$ Pa, the cathode potential is -24.8 kV, which gives the initial electric field ~24.8 kV/cm.

Figure 8 shows the time evolution of the electron density, the potential and the electric field. Figure 9 shows the time evolution of the VDF at the distance of 5 and 7.5 mm from the cathode and Figure 10 shows the electron density in phase space obtained at four different times.

As it follows from the results of simulation, the emitted electrons generate rather dense quasi-neutral plasma at some distance from the cathode. At this distance, the emitted electrons gain the energy $w = I/2$. Once the plasma density in the vicinity of the cathode reaches the value ~$10^{18}$ m$^{-3}$, it starts screening the applied electric field. Then, the highest electric field is obtained between the cathode and the plasma (cathode sheath) and between the plasma and the anode (see Figure 8(c)). These electric fields are homogeneous due to the planar geometry. The phase space shown in Figure 10 allows one to conclude that the emitted electrons form the beam of runaway electrons. After the generation of dense plasma in the vicinity of the cathode, this beam has the energy equal to the cathode sheath voltage. This beam dissipates its energy while it propagates through the quasi-neutral plasma since the electric field in this plasma is much smaller than the electric field obtained in the cathode sheath (Figure 8(c)). However, when the beam enters the space between the plasma and the anode, it starts gaining the energy again. One can conclude from Figure 10 that there are the runaway electrons between the plasma and the anode.

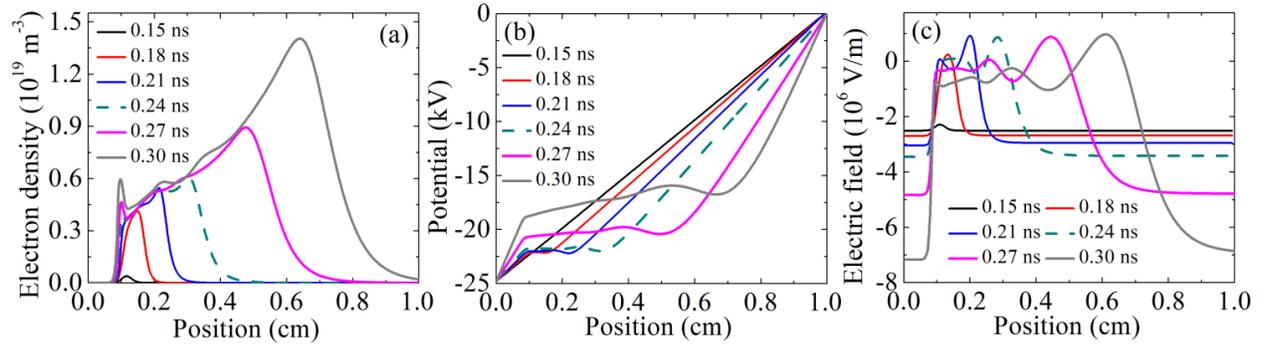

Figure 8. Spatial profiles of (a) the electron density, (b) electrostatic potential, and (c) electric field obtained at different times for the electron-emission driven discharge.

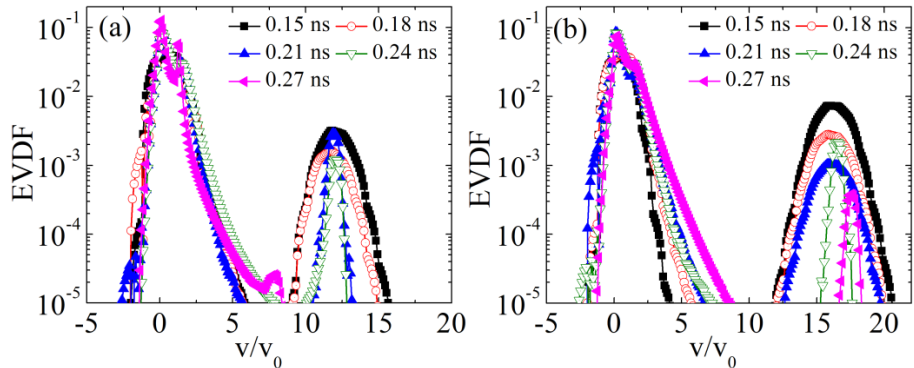

Figure 9. Electron velocity distribution function (in s/m$^4$) obtained at different times at the distance of (a) 5 mm and (b) 7.5 mm from the cathode for the electron emission-driven discharge.

Figure 8 allows one to conclude that there is a fast ionization wave propagating from the cathode to the anode. This wave propagates toward the anode due to multiplication of background electrons that are present in front of the wave. These electrons are created by the beam of runaway electrons. The density of the background electrons is small (~$10^{15}$ m$^{-3}$) and not sufficient to screen the applied electric field (Figure 8(c)). Therefore, they gain high energy from the electric field and support the propagation of the fast ionization wave toward the anode.

Figure 8(c) shows that the electric field between the wave and the anode increases with time. This is obtained because the distance between the front of the ionization wave and the anode decreases. The increase of the electric field in turn leads to an increase of the ionization rate. Since the ionization rate defines the speed of the ionization wave propagation, one can conclude that the wave speed increases with time.

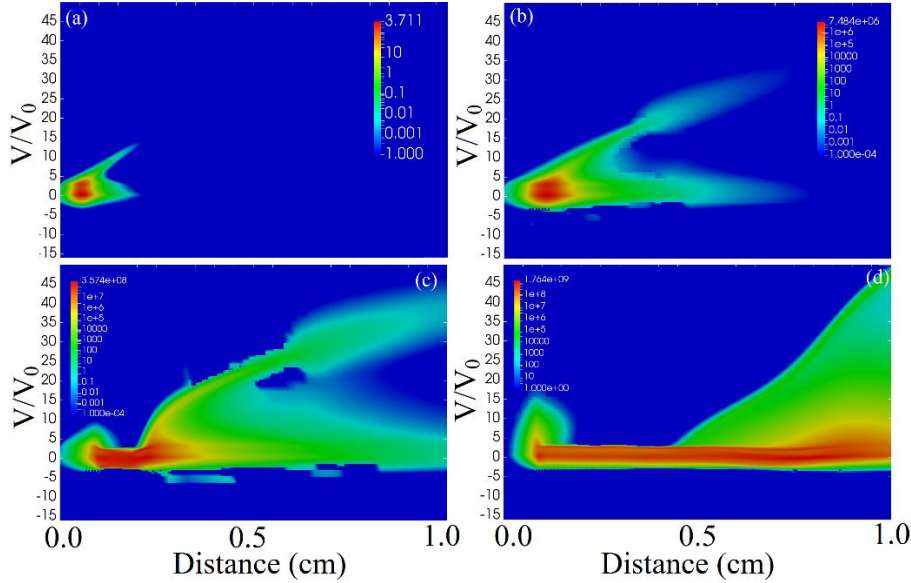

Figure 10. Electron density in phase space obtained at different times: (a) 0.15 ns, (b) 0.21 ns, (c) 0.24 ns, and (d) 0.27 ns for the electron emission-driven discharge.

Another interesting effect obtained in the present simulation is the generation of the so-called "anomalous" electrons. The presence of such electrons in the cathode-anode gap follows from the detailed analysis of the results shown in Figure 9(b) and Figure 8(b). One can conclude from Figure 8 that the average velocity of the ionization front propagation is $v_{front} \sim 2.3 \times 10^7$ m/s. The maximum electron energy $w \sim 20$ keV is obtained at $t = 0.27$ ns at the distance of 7.5 mm from the cathode. As it follows from Figure 8(b), the electron emitted from the cathode propagates the potential difference $\Delta\varphi \sim 10$ kV, i.e. $e\Delta\varphi < w$. However, considering that the energy of electron "sitting" on the wave is defined as $\frac{1}{2}m_e(v_e + v_{front})^2$, one defines the electron energy $w' \sim 19.5$ keV which is in a rather good agreement with the value obtained in our simulations.

The generation of anomalous electrons in nanosecond discharges was predicted long ago by Askaryan[14] and recently demonstrated by Particle-in-Cell simulations.[15] Following this mechanism, the runaway electrons presented between the anode and the "head" of the ionization front are pushed by the front (so-called wave riding). The electrons obtain an additional velocity from the moving front if its velocity is comparable with the velocity of runaway electrons. The velocity of runaway electrons is $>10^7$ m/s. As discussed in Ref. [15], such high velocities of the ionization front can be obtained in discharges driven by electron emission from cathode because in this case the density of runaway electrons is high enough to generate rather dense background in front of the ionization wave. The number density of the background electrons defines the velocity of the ionization wave.[16]

## CONCLUSIONS

We have developed a Boltzmann-Fokker-Planck kinetic solver with adaptive Cartesian mesh in 1d2v phase space using spherical coordinates in velocity space to simplify calculations of collisions in both Lorentz and Fokker-Planck forms. We have illustrated key features of the solver for elastic scattering, acceleration, continuous energy loss in inelastic collisions, electron runaway, and multiplication in spatially inhomogeneous electric fields. Having coupled the kinetic solver with Poisson solver, we studied gas breakdown dynamics in the presence of runaway electrons. We have identified limitations of the non-split-phase-space-mesh method for strongly anisotropic VDFs, where fine mesh in the μ-direction forces unnecessary mesh refinement in the *x* and *v*-directions. Future work will include implementation of discrete model for inelastic collisions at energies close to the electron excitation threshold of neutrals and further development of the AMPS methodology for kinetic solvers.

## ACKNOWLEDGMENTS

This work is supported by the DOE SBIR Project DE-SC0015746, by the US Department of Energy Office of Fusion Energy Science Contract DE-SC0001939, and the NSF EPSCoR project OIA-1655280 "Connecting the Plasma Universe to Plasma Technology in AL: The Science and Technology of Low-Temperature Plasma".

## REFERENCES

[1] S. Zabelok, R. Arslanbekov, and V. Kolobov, J. Comput. Phys. **303**, 455 (2015).
[2] M. Palmroth et al., Living Rev. Comput. Astrophys. **4**, 1 (2018); https://doi.org/10.1007/s41115-018-0003-2.
[3] R. R. Arslanbekov, V. I. Kolobov, and A. A. Frolova, Phys. Rev. E **88**, 063301 (2013)
[4] V. Kolobov, R. Arslanbekov and D. Levko, Kinetic solvers with Adaptive Mesh in Phase Space for Low-Temperature Plasmas, https://arxiv.org/abs/1809.05061.
[5] D J Strickland et al., J. Geophys. Res. **81**, 2755 (1976).
[6] http://basilisk.fr
[7] R. J. Danos, J. D. Fiege, and A. Shalchi, Astrophys. Journal **772**, 35 (2013).
[8] Yu. I. Fedorov and B. A. Shakhov, Kinematics and Physics of Celestial Bodies **34**, 107 (2018).
[9] K Stamnes, J. Geophys. Res. **86**, 2405 (1981).
[10] Yu. P Raizer and M. N. Shneider, Fizika Plasmy **15**, 318 (1989).
[11] V.I. Kolobov and L.D. Tsendin, Phys. Rev. A **46**, 7837 (1992).
[12] L. D. Tsendin and D. S. Nikandrov, Plasma Source Sci. Technol. **18,** 035007 (2009).
[13] Yu. D. Korolev, G. A. Mesyats, *Physics of Pulse Breakdown in Gases* (Nauka, Moscow, 1990).
[14] G. A. Askaryan, JETP Lett. **2,** 113 (1965).
[15] D. Levko, S. Yatom, and Ya. E. Krasik, J. Appl. Phys. **123,** 083303 (2018).
[16] S. I. Yakovlenko, Technical Physics **49,** 1150 (2004).